\documentclass[conference]{IEEEtran}
\ifCLASSINFOpdf
\else
\fi

\usepackage{math}
\usepackage{subcaption}	
\usepackage{graphicx}
\usepackage{adjustbox}
\usepackage{longtable}

\def\etal{{\it et al.~}}

\newcommand{\mc}[1]{\mathcal{#1}}

\begin{document}
%
\title{Forecasting of commercial sales \\with large scale Gaussian Processes}

\author{\IEEEauthorblockN{Rodrigo Rivera}
\IEEEauthorblockA{School of Computer Science,\\
Higher School of Economics\\
Email: rriverakastro@edu.hse.ru}
\and
\IEEEauthorblockN{Evgeny Burnaev}
\IEEEauthorblockA{Skolkovo Institute of Science and Technology,\\
Institute for Information Transmission Problems\\
Email: e.burnaev@skoltech.ru}
}


%


\maketitle

\begin{abstract}
This paper argues that there has not been enough discussion in the field of applications of Gaussian Process for the fast moving consumer goods industry. Yet, this technique can be important as it e.g., can provide automatic feature relevance determination and the posterior mean can unlock insights on the data. Significant challenges are the large size and high dimensionality of commercial data at a point of sale. The study reviews approaches in the Gaussian Processes modeling for large data sets, evaluates their performance on commercial sales and shows value of this type of models as a decision-making tool for management.
\end{abstract}
Keywords: Gaussian Processes, demand forecasting, retail, fast moving consumer goods

%
\IEEEpeerreviewmaketitle

\section{Introduction}

This study seeks to contribute to a better understanding in the industry to the field of forecasting with Gaussian Processes (\textit{GPs}) with a focus on large data sets of commercial sales data for consumer goods in the fast moving consumer goods (\textit{FMCG}) industry.
Whereas there has been a wide interest in academia to explore Gaussian Processes for small data sets and on developing methods to fit large data into Gaussian Processes, there has not been enough diffusion for the commercial practitioner. Yet, \textit{GPs} have found success on various applications. They are in use on a wide range of fields and use cases ranging from surrogate modeling, experiment design, mining and geo-spatial data to battery health \cite{Banerjee2008, 2017arXiv170305687R,burnaev2013,burnaev2015,burnaev2017,burnaev20171,burnaev2016,burnaev20161, burnaev20162, burnaev12015, burnaev20151}. On techniques for demand forecasting in the \textit{FMCG} sector, there has been little academic research and not enough efforts to expose practitioners to them. Other forecasting techniques for consumer demand are widely used in industry and academia. For example, approaches using time series analysis, support vector machines (SVM), neural networks and splines \cite{Lu2012, Ding2014}. The industry has not been completely oblivious to employing \textit{GPs} to forecast demand, for example, in electricity \cite{Blum2013, Samarasinghe2012}, windmills \cite{Chen2013}, e-commerce \cite{Huang2015}, water \cite{Wang2014} and tourism \cite{Wu2012, Claveria2017}. Similarly, the consumer staples industry has made an effort to apply principled approaches to demand forecasting. One case was for tobacco products using seasonal time series decomposition and support vector machines to predict aggregations either of total industry demand \cite{Tan2015} or of an individual producer \cite{Ding2014}. GPs is a popular non-parametric model for regression and classification tasks leveraging the power of the underlying Bayesian inference framework \cite{Rasmussen2006, 9780262182539}. One shortcoming is that the computational performance of \textit{GPs} deteriorates fast for data sets with more than thousands of observations \cite{Lawrence2016} such as those found in commercial settings \cite{Seeger2004}. The objective of this research is to provide an overview for the practitioner of the theory and implementations of \textit{GPs} in a big data setting using large commercial data sets as examples. The study argues that \textit{GP} is a viable tool for sales forecasting and demand prediction in the consumer staples industry at a point of sale (POS). Recent advances with a focus on big data make them robust methods for multidimensional data \cite{Lawrence2016, Zhang2017, DBLP:conf/uai/2013, 2014arXiv1402.1389G, daviesphd2014}. Similarly, their flexibility as predictors qualify them for the particular characteristics of demand prediction for consumer products. This study evaluates and describes the characteristics of \textit{GPs} for big data applied to commercial data sets of the \textit{FMCG} industry and assesses related software libraries. For this purpose, it poses the following questions: (1) What is the state of the art in academic research of Gaussian Processes for big data? (2) Which implementations of Gaussian Processes are currently available? (3) How do \textit{GPs} perform in practical tasks in the \textit{FMCG} industry?
\subsection{Significance \& Innovation}
Demand planners continue leveraging traditional forecasting techniques \cite{Chase2013}. From an academic perspective, studies on sales forecasting either do not cover \textit{FMCG} or use regression techniques considered to be special cases of \textit{GPs} \cite{9780262182539}. While valuable, these techniques do not benefit from the advantages of a probabilistic approach such as \textit{GPs}. Moreover, there has been a limited number of academic studies in the area of \textit{GPs} combining the disciplines of management sciences and computer sciences \cite{Dew2016, Dew2016a}. An additional difficulty to assess the suitability of \textit{GPs} for big commercial data is the popularity in the literature of small and synthetic data sets for benchmarks. This study innovates by combining a multidisciplinary approach on how \textit{GPs} are evaluated from a theoretical and practical perspective and by bringing closer the \textit{GP} theory and application to the industrial practitioner. Apart from a methodological innovation, this research seeks to contribute to a better understanding of Gaussian Process regression (GPR) for large commercial data in the \textit{FMCG} industry.
\section{Related Work}
There is vast literature on \textit{GPs}, for example Rasmussen \& Williams \cite{9780262182539}. Summarized, a Gaussian process is a stochastic process  $h(x)$ such that any sub-collection of random variables $h(x_1),h(x_2),\ldots,h(x_n)$ for any finite subset of elements $x_1,\ldots,x_n \in X \subset \mathbb{R}^d$  follows a multivariate Gaussian distribution. The key characteristic of a \textit{GP} is that it is fully specified by a mean function $m(x)$ and a covariance function $k(x,x')$. Thus,
\begin{equation}\label{eq:gpdef}
h(x)\sim \mc{GP}(m(x),k(x,x')).
\end{equation}
The \textit{GP} generalizes the Gaussian distribution going from vectors to functions. Further, it allows for efficient computation since it only needs values at a finite number of points. Moreover, the smoothness of a \textit{GP} is controlled by the covariance function $k(x,x')$. For Gaussian Process regression (GPR), the prior is assumed to be a \textit{GP}, see definition \eqref{eq:gpdef}, with a zero mean.
Suppose there are $n$ points $x_i, i=1,2,\ldots,n$ and corresponding labels $\bar{y} = (y_1,\ldots,y_n)$. For a new point $x_*$, $y_* = f(x_*)$ should be predicted. By assuming that the regression function $f$ is a Gaussian process, it is obtained that the joint distribution of the training outputs $\bar{y}$ and the test outputs $y_*$, both vectors, at any finite number of points is a Gaussian
\begin{equation}
\begin{pmatrix} \bar{y} \\
y_* \end{pmatrix} \sim \mc{N}\Biggl(0,\begin{pmatrix} K & K'_*\\
 K_* & K_{**} \end{pmatrix}\Biggr).
\end{equation}
This is called the joint prior, where the mean equals zero, $K$ is the covariance matrix for the training points $x_i$, $K_*$ is a covariance vector between the test $x_*$ and the training points $x_i$, and $K_{**}$ is the inherent measurement noise. To get the posterior distribution over functions, it is necessary to condition this joint Gaussian prior distribution on the observations as following
\begin{equation}
p(y_*|\bar{y}) \sim \mc{N}(K_*K^{-1}y,K_{**}-K_*K^{-1}K'_*).
\end{equation}
\subsection{Advantages \& Disadvantages}
Some of the advantages attributed to \textit{GPs} are according to Do \cite{do2007}: they enable the possibility to quantify uncertainty in predictions resulting not only from the intrinsic noise in the problem but also from the errors in the parameter estimation procedure. Further, many methods for model selection and hyperparameter selection in Bayesian frameworks can be applied. Moreover, GPR is non-parametric. Thus, it can model any arbitrary function based on the input points. In addition, they present a natural way to introduce kernels into the regression modeling framework.
At the same time, their main disadvantages are the computation and storage costs, dominated by the inversion of $K$, as well as the difficulty to understand conceptually the theory behind. Moreover, they do not deal well with discontinuities such as those found in financial crises, phosphorylation, collisions, edges in images, etc. In addition, Lawrence claims that the popular squared exponential covariance (RBF) often used in the literature is too smooth for practical problems \cite{do2007, Lawrence2016}.
\subsection{Gaussian Processes for Big Data}\label{ch2:gpbigdata}
The largest limitations of using \textit{GPs} in practice are the storage $\mc{O}(n^2)$ and computational $\mc{O}(n^3)$ requirements for $n$ data points. Nevertheless, the strengths of \textit{GP} such as its flexibility, its conceptual simplicity and desirable properties make it highly attractive to address regression tasks. For this reason, there has been an active interest in the research community to extend \textit{GPs} to large data sets.
This study organizes \textit{GPs} methods for big data around four themes: (1) Sparse approximations, (2) Low-rank approximations, (3) Local approximations, (4) Parallelization. Other authors such as Hoang \cite{hoang2014} and Rasmussen \& Williams \cite{9780262182539} follow other classifications.
\subsection{Sparse approximations}\label{ch2:gpbigdata-sa}
The general idea of a sparse approximation is to have a sample covariance matrix that is sparser than its original. The inversion of the sparse matrix is less computationally expensive than the inversion of a non-sparse matrix of the same size. This category has a very rich list of heterogeneous approaches. Furrer \etal \cite{furrer2006} and Kaufman \etal \cite{Kaufman08covariancetapering} apply a covariance tapering technique. On the other hand, Gneiting \cite{CIS-180533} uses a compactly supported covariance function. Lindgren \etal \cite{Lindgren_2011} use the Gaussian Markov approximation of a GP. The work of Grigorievskiy \etal \cite{2016arXiv161008035G} for temporal \textit{GPs} with a 1-dimensional input space uses the sparseness property of precision matrices in a Markovian process to scale computationally to $\mc{O}(b^3n)$ where $b$ is the size of the matrix block. On a similar direction, Gilboa \etal \cite{gilboa2013} also use a Markov process under the umbrella of the projection pursuit method for structured \textit{GPs}. A \textit{GP} can be considered structured, if its marginals contain exploitable structure enabling a reduction in computational complexity. 
Another family of approximations is based on approximate matrix-vector-multiplications (MVMs). Some of these methods have been reviewed by Quinonero-Candela \etal \cite{4798}. Local mixtures of \textit{GPs} have been used by Urtasun \& Darrell \cite{conf/cvpr/UrtasunD08} for efficient modeling of human poses. Gal \etal \cite{2014arXiv1402.1389G} use a re-parametrization of variational inference. On a different direction, Davies \& Ghahramani \cite{2014arXiv1402.4293D} propose focusing on the kernel and using matrix-free methods. They do not require the full Gram matrix $K$, only the ability to calculate $Kv$ for any arbitrary $v$ \cite{Titsias09variationallearning}.
\subsection{Low-rank approximations}\label{ch2:gpbigdata-lra}
Historically, most of the work on \textit{GPs} for large datasets has been focused on this area according to Hoang \etal \cite{2016arXiv1611.06080}. The low-rank approximate representation of the full-rank \textit{GP} (FGP) is an alternative family for sparse \textit{GP} regression. Here, the idea is to build a low rank approximation of the covariance matrix based around so-called 'inducing variables'. Examples of this are the works of Csato \& Opper \cite{csatoopper2002}, Seeger \etal \cite{Seeger03fastforward}, Quinonero-Candela \& Rasmussen \cite{2745} and Titsias \cite{Titsias09variationallearning}.
Low-rank approximations introduce latent variables and assume a certain independence conditioned on the latent variables leading to a computational complexity of $\mc{O}(nm^2)$ and storage demands of $\mc{O}(nm)$ with $m$ as the parameter governing the number of selected inducing variables \cite{Seeger03fastforward, NIPS2005_2857}. For example Hensman \etal \cite{DBLP:conf/uai/2013} do this with variational inference. 
According to Low \etal \cite{2014arXiv1411.4510L}, low-rank methods are well-suited for modeling slowly-varying functions that are largely correlated, and where it is possible to use all the data for predictions. The downside is that they require a relatively high rank to capture small correlations with high fidelity; this makes them lose attractiveness. Authors such as Low \etal \cite{2014arXiv1411.4510L} try to overcome this by leveraging the dual computation of complementing a low-rank approximate representation of the full-rank \textit{GP} based on a support set of inputs with a Markov approximation of the resulting residual process.
Hoang \etal \cite{2016arXiv1611.06080} see here two large subgroups. On one side the \textit{distributed} approach, where it is sought to reduce the training time with all the data by a factor close to the number of machines. Some examples here are the works of Chen \etal \cite{2014arXiv1408.2060C}, Hoang \etal \cite{2016arXiv1611.06080, hoang206} and Low \etal \cite{pmlr-v37-hoang15}. 
The second subgroup is the \textit{stochastic} implementation. The main idea behind is twofold. First, to train with a small, randomly sampled subset of data in constant time per iteration of stochastic gradient ascent update. Second, it is to achieve asymptotic convergence to their predictive distributions. One example is Hensman \etal \cite{DBLP:conf/uai/2013}. 
Das \etal \cite{DBLP:journals/corr/DasRS15} follow a different approach and opt for an empirical method of first using a bootstrapped data set to develop a \textit{GP} and then to bag the models to produce the regression estimate. Finally, Hoang \etal \cite{2016arXiv1611.06080} argue that there is a less well-explored class exploiting sparsity in the spectral representation of a \textit{GP} kernel.
\subsection{Local approximations}\label{ch2:gpbigdata-la}
In this approach the general idea is to partition the data set into separate groups. Exponents of this concept are Snelson \& Ghahramani \cite{Snelson07localand} and Urtasun \& Darrell. Low \etal \cite{2014arXiv1411.4510L} mention that this family comprises ideas around localized regression, covariance tapering methods and compactly supported covariance functions. Their strength is that they address the shortcomings of low-rank approximations; they can model rapidly-varying functions with small correlations. As they only use local data for predictions, they end up performing poorly in input regions with little data.
One class of local approximations partitions the input domain into a set of local regions and assume an independent \textit{GP} regression model within each region. The resulting sample covariance matrix is a block diagonal matrix of local sample covariance matrices. Park \& Apley \cite{2017arXiv170106655P} highlight that inverting the block diagonal matrix is much cheaper computationally. Such local approximation approaches have many advantages. By their local nature, they adapt better to local and non-stationary data features and independent local approximation models can be computed in parallel to reduce total computation time. Their major weakness is that two local models for two neighboring local regions produce different predictions at the boundary between the regions, resulting in boundary discontinuity for the regression predictive function. This boundary discontinuity implies greater degradation in prediction accuracy, particularly around the boundaries of the local regions \cite{JMLR:v17:15-327}.
The discontinuity issue has been addressed in different ways. The most popular approach is to smooth out some of the discontinuity by using some weighted average across the local models or across multiple sets of local models via a Dirichlet mixture (Rasmussen \& Ghahramani \cite{Rasmussen01infinitemixtures}), or via a treed mixture (Gramacy \& Lee \cite{Arxiv0710.4536}), or with a Bayesian model averaging (Tresp \cite{Tresp_2000}, Chen \& Ren \cite{chenren2009} and Deisenroth \& Ng \cite{2015arXiv150202843D}), or with locally weighted projections (Nguyen-Tuong \etal \cite{Nguyen_Tuong_2009}).
Other related approaches use an additive covariance function consisting of a global covariance and a local covariance. Examples are the works of Snelson \& Ghahramani \cite{Snelson07localand} and Vanhatalo \& Vehtari \cite{2012arXiv1206.3290V}. Another alternative is to construct a local model for each testing point (Gramacy \& Apley \cite{2013arXiv1303.0383G}), or to use a local partition but constrain the local models for continuity. This last proposal is reflected in the works of Park \& Huang \cite{JMLR:v17:15-327} and Park \etal \cite{Park:2011:DDA:1953048.2021054}.
\subsection{Parallelization}\label{ch2:gpbigdata-p}
More than a technique to deal with data, parallelization is rather used jointly with other techniques described above. Sparse \textit{GPs} can for example make use of existing paradigms in distributed computing such as Apache Spark \cite{Zaharia:2010:SCC:1863103.1863113} or Map-Reduce \cite{Dean:2004:MSD:1251254.1251264}. 
In the work of Gal \etal \cite{2014arXiv1402.1389G}, they re-parametrize variational inference for sparse GPR, see section \ref{ch2:gpbigdata-lra}, to re-formulate the evidence lower bound in a Map-Reduce setting. Low \etal \cite{2014arXiv1411.4510L} propose a \textit{low-rank-cum-Markov approximation} (LMA) of the full \textit{GP}. This work falls also into the family of methods discussed in section \ref{ch2:gpbigdata-la}. Similarly, Zhang \& Williamson \cite{Zhang2017} propose a parallel local method under the \textit{manteau} of 'embarrassingly parallel' algorithms, where the global communication occurs only after the local computation is complete. On the same vein, Deisenroth \& Ng \cite{2015arXiv150202843D} exploit the fact that a single \textit{GP} can be split into $q$ independent problems, whose parameters can be inferred independently (and parallel) of each other in a weighted product-of-experts model \cite{deisenroth20151, deisenroth2015}. An example of a sparse approximation, see section \ref{ch2:gpbigdata-sa}, with a parallel component is in the work of Grigorievskiy \etal \cite{2016arXiv161008035G}.	
Another area of research lies on the parallelization of \textit{GPs} through distributed computing or a combination of CPU and GPU as shown by Gramacy \etal \cite{2013arXiv1310.5182G}. They show that the combined effects of approximation and massive parallelization can be applied to \textit{GPs}. However, this requires careful work.
\section{Advice for the practitioner}\label{ch2:aftp}
Inspired by the work of Davies \cite{daviesphd2014}, this study suggests following best practices for the practitioner. In kernel-related methods, the first and most important step is to select an appropriate kernel. It should be remembered that the kernel encodes all the domain knowledge needed. Many properties can be embedded into the kernel such as smoothness, periodicity, linearity, dependence between dimensions, etc. The literature on kernels and kernel choice is vast. The reader is advised to choose a covariance function that reflects the known properties of the underlying process as much as possible.
On a second step, it is necessary to choose the ideal approximation. There is not a 'one-size fits all' approximation akin to the choice of kernel.
The third step is to determine the required level of accuracy for the posterior. This is closely related to the choice of kernel and of approximation. Often, problems do not require an accuracy that justifies a computational cost of $\mc{O}(n^3)$. The practitioner should evaluate the problem at hand and determine the desired trade-off between computational speed and accuracy.
Frequently, hyperparameters need to be optimized. Thus, the fourth step is choosing an ideal optimization. A popular choice for hyperparameter optimization is the log-likelihood.
Finally, probably the most important step is data preparation. Often, not all data is necessary for a problem. It is desirable to work with a subset of the original data set.
\section{Implementations of Gaussian Processes}\label{ch3:intro}
To strengthen the notion that \textit{GPs} are both popular and used in practice, table \ref{tab:gppopularity}\footnote{ARMA \& Arima used interchangeably. Splines as spline regression} portrays a comparison between \textit{GPs} and other forecasting techniques on diverse Internet websites. For example \textit{GP} regression is frequently listed as part of a repository description in Github\footnote{http://www.github.com}. Similar results were obtained from comparing the level of popularity since 2004 using Google Trends\footnote{http://trends.google.com} and from counting the number of threads under the respective tag at CrossValidated (CV)\footnote{https://stats.stackexchange.com/} as well as at Reddit\footnote{http://www.reddit.com} respectively. 
\begin{table}[ht]
	\caption{Popularity of regression techniques on public forums by ranking (best $=1$)} \label{tab:gppopularity} 
	\centering
	\begin{adjustbox}{max size={\textwidth}{\textheight}}
	\small
  \begin{tabular}{| p{2.6cm} | p{1cm} | p{1.2cm} | p{0.6cm} | p{1cm} |} \hline
  Technique & Github & GTrends & CV & Reddit \\ \hline
  Bayesian linear regression & 5 & 6 & 5 & 3 \\ \hline
  Kernel ridge regression & 7 & 5 & 3 & 7 \\ \hline
  Splines & 6 & 7 & 4 & 6 \\ \hline
  NN regression & 1 & 2 & 5 & 1 \\ \hline
  SVM regression & 3 & 3 & 5 & 2 \\ \hline
  ARMA & 2 & 1 & 1 & 4 \\ \hline
  GP regression & 4 & 4 & 2 & 5 \\ \hline
  \end{tabular}
	\end{adjustbox}
\end{table}
The results show that \textit{GPs}, although still obscure, have a stronger presence in public forums than 'classical' regression methods such as Splines and Kernel ridge regression (KRR).
Similarly, it is of relevance for the practitioner to have a library of choice for prediction with \textit{GPs}. This study compares popular implementations of \textit{GP} regression under a criteria inspired by Golge \cite{erogol2016} and De \etal \cite{De2016} considering soft factors such as an active community and hard ones such as computational performance. 
\section{Gaussian process libraries}\label{ch3:gplibs}
To assess different implementations of GPR, this research assigns up to three points, (1) acceptable, (2) good and (3) very good, to each of the following aspects.
\paragraph{Community} A large community is a sign of the long-term viability and commitment on an open source software project.
\paragraph{Documentation} Extensive documentation shows commitment on the library and an active community.
\paragraph{Stability} Often, new libraries undergo massive changes. For the practitioner, this complicates selling the technology within the organization.
\paragraph{Run-time performance} \textit{GPs} have a reputation for being slow. Newer libraries try to speed this up through GPU support, distributed computation \& optimized implementations.
\paragraph{Flexibility} Experimental libraries offer often a higher degree of flexibility and experimentation. This can be helpful to understand \textit{GPs} on a deeper level or to implement a customized optimization, kernel or approximation method.
\paragraph{Development} It is desirable that the latest theoretical advances are reflected in the library of choice.
\paragraph{Examples} Good and comprehensive examples are an important part of the familiarization and education process.
\paragraph{Object Oriented} Libraries with object-oriented APIs are desired for deployment in production environment.
\paragraph{Test coverage} Often, organizations set guidelines on a minimum test coverage to be fulfilled. For the practitioner seeking to introduce \textit{GPs}, test coverage is essential.
For the purpose of this analysis, a set of \textit{GP} regression libraries was selected in table \ref{tab:gpimpoverview}.
Sci-Kit Learn \cite{FABIANPEDREGOSA2011}, GPy \cite{gpy2014}, GPflow \cite{De2016}, Edward \cite{Tran2017}, George \cite{2017arXiv170309710F, hodlr}, pyGP \cite{Neumann:2015:PPL:2789272.2912082}, pyMC3 \cite{Salvatier2016}. Implementations in other programming languages were also assessed. In R, GPfit \cite{gpfit2013} and gptk\footnote{https://github.com/SheffieldML/gptk} are popular. However, they have not been updated in the last years.
Julia, another language used by statisticians, does not appear to have an established \textit{GP} library. This study found juliaGP\footnote{https://github.com/st--/juliagp}, MLKernels\footnote{https://github.com/trthatcher/MLKernels.jl} and two libraries named GaussianProcesses.jl\footnote{https://github.com/linxihui/GaussianProcess.jl}\footnote{https://github.com/STOR-i/GaussianProcesses.jl}. As of this writing, none of them is either mature or active.
Stan \cite{Carpenter_stan}, another popular programming language among Bayesian statisticians has a strong community, extensive documentation and an API for Python, PyStan. However, to narrow down the scope of this study, only primarily Python libraries are considered. This is not the first study evaluating \textit{GP} software empirically. Erickson \etal \cite{Erickson_2016} also compared multiple \textit{GP} implementations. Their conclusion was that Scikit-learn and GPy are among the best options. However, they warn that two different implementations can give significantly different results on the same data set and therefore advice to select a library on a by-case basis. The libraries listed below were chosen for further evaluation with table \ref{tab:gpimpoverview} summarizing the assessment.
\paragraph{Scikit-learn}\label{ch3:scikit} It includes implementations of almost all major machine learning methods. The community is large and active. Due to its popularity, there is extensive documentation and examples. The library is stable, with clear release cycles and support for legacy versions. On the other hand, it is not optimized for GPU. Similarly, it is less-flexible than other libraries evaluated, object-oriented and has extensive test coverage. 
\paragraph{GPy} The most popular library in the \textit{GP} community among libraries exclusively dedicated to \textit{GP} modeling. The library has a strong community and multiple examples. The documentation is good. Additionally, it is also object-oriented This library can be seen as a compromise between industrial requirements and academic needs. Downsides are limited GPU functionality and insufficient test coverage.
\paragraph{GPflow}\label{ch3:gpflow} The focus is primarily on experimentation without sacrificing test coverage and speed. It uses full GPU acceleration and variational inference as the default approximation method. The downside is that the library is not as well-known as others. The community is small but active. Similarly, although the standard documentation is comprehensive, the amount of available examples is rather limited.
\paragraph{Edward} An alternative to Stan in Python. The focus is on Bayesian probabilistic modeling and on academic experimentation. Akin to GPflow, the standard documentation is optimal with limited examples outside the official material. It has a small and active community. Given that it leverages TensorFlow \cite{tensorflow2016}, it can speed up computations with GPU.
\paragraph{George} A library with a focus on fast \textit{GP} regression. Compared to the previous libraries, it is less popular. This is reflected on the sparse official documentation and on the lack of examples. In addition, development has stalled and the community support seems limited. Moreover, it does not offer hardware acceleration and the test coverage is basic. 
\paragraph{pyGP} It has not been updated in many years and the community is dormant. It lacks examples and the documentation is basic.
\paragraph{pyMC3}\label{ch3:pymc3} Akin to Stan and Edward, pyMC3 is a probabilistic programming library. It is well-known in both academia and industry. The community is large, the documentation comprehensive and many examples are easily found. Given its level of industrial adoption, the library is stable and has well-known development cycles. PyMC3 is an interesting option for the industrial practitioner interested in Bayesian inference on a production-ready environment.
\begin{table}[ht]
	\caption{Libraries for Gaussian Process Regression by scores (3: best)} \label{tab:gpimpoverview} 
	\centering
	\begin{adjustbox}{max size={\textwidth}{\textheight}}
	\small
  \begin{tabular}{| p{0.9cm} | p{0.48cm} | p{0.48cm} | p{0.48cm} | p{0.48cm} | p{0.48cm} | p{0.48cm} | p{0.28cm} | p{0.28cm} | p{0.28cm} |} \hline
  Library & Com & Doc & Stab & Perf & Flex & Dev & EX & OO & TC \\ \hline
  Scikit-learn & 3 & 3 & 3 & 2 & 1 & 3 & 3 & 3 & 3 \\ \hline
  GPy & 3 & 2 & 3 & 2 & 2 & 2 & 2 & 3 & 2 \\ \hline
  GPflow & 2 & 2 & 2 & 3 & 3 & 3 & 2 & 3 & 3 \\ \hline
  Edward & 2 & 2 & 2 & 3 & 3 & 3 & 2 & 3 & 2 \\ \hline
  George & 1 & 1 & 2 & 2 & 2 & 1 & 1 & 2 & 2 \\ \hline
  py\textit{GP} & 1 & 1 & 2 & 1 & 2 & 1 & 1 & 2 & 1 \\ \hline
  pyMC3 & 3 & 3 & 3 & 3 & 1 & 3 & 3 & 3 & 3 \\ \hline
  \end{tabular}
	\end{adjustbox}
	\bigskip
	\\Com = Community, Doc = Documentation, Stab = Stability, Perf = Performance, Flex = Flexibility, Dev = Development, EX = Examples, OO = Object Oriented, TC = Test Coverage
\end{table}
Based on the scores from this assessment, Scikit-learn, pyMC3 and GPflow are the stronger choices for \textit{GPs} in Python.
\section{Data sets analysis}\label{ch5}
The Grupo Bimbo Inventory Demand Kaggle\footnote{https://www.kaggle.com/c/grupo-bimbo-inventory-demand} competition data and a data set provided by a Fortune 500\footnote{http://fortune.com/fortune500/} manufacturer of consumer goods were analyzed\footnote{https://github.com/rodrigorivera/forecastingcommercial}. Both share similar characteristics as seen in table \ref{tab:datasetspoverview} such as large number of observations, presence of categorical features and industry. Similarly, the objective is to predict demand by stock keeping unit (SKU), a product, at the POS. As a first step, an exploratory analysis was conducted.
\begin{table}[ht]
	\caption{Overview of commercial data sets} \label{tab:datasetspoverview} 
	\centering
	\begin{adjustbox}{max size={\textwidth}{\textheight}}
	\small
  \begin{tabular}{| p{2.1cm} | p{1.5cm} | p{1.5cm} | p{1.8cm} |} \hline
  Data set & Numerical features & Categorical features & Observations \\ \hline
  Bimbo original & 6 & 5 & 7.4M \\ \hline
  Fortune 500 original & 4 & 6 & 10.6M \\ \hline
  Bimbo post-process & 147 & 0 & 9.05M \\ \hline
  Fortune 500 post-process & 112 & 0 & 10.35M \\ \hline
  \end{tabular}
	\end{adjustbox}
\end{table}
Based on this, it was decided to process both data sets as following: (1) Group POS along broader trade categories (e.g., supermarkets, universities) and remove stop words and generic details, (2) Normalize attributes, one-hot-encode categorical features and remove superfluous observations, (3) Split training set and use only one week of data, (4) Log transform the target variable (5) For other weeks, calculate sample out-of-mean features, (6) Group and join categorical features with other attributes and generate descriptive statistics (mean, median, standard deviation, sum) related to demand for each categorical feature, (7) Drop observations with null values for \textit{GP} methods and retain for XGBoost \cite{2016arXiv160302754C}, (8) Use XGboost to identify relevant features. The evaluation of both data sets was a time consuming effort. Regardless of the level of domain knowledge, it was necessary to invest in exploratory analysis. Initial approaches for fitting the data did not yield good results and due to its characteristics, iterations came at a significant time cost.
\subsection{Bimbo data set}
There have been previous analyses on this data set, for example by Kosar \etal \cite{Kosar2016}. Important to highlight, the test set does not contain numerical variables. Thus, demand has to be predicted using categorical features. Additional files with attributes related to POS and their location were also made available. There is a relation between sales and returns. Locations with low levels of sales have high level of returns. In addition, evaluating the data showed irregularities in its collection. After pre-processing, the resulting data set contained 147 features. The five most relevant based on their XGboost gain were (1) Average returns by POS type, (2) Average sales by POS type and depot, (3) Average log demand by POS type and depot, (4) Average log demand by POS type, (5) Average log demand by POS type and city.
\subsection{Fortune 500 company data set}
Data generated at POS for 28 weeks in 2016 was provided containing attributes such as sales volume per SKU, type of retail outlet, location, stock levels, trade marketing activities \& more.
\begin{figure}
\includegraphics[width=0.5\textwidth]{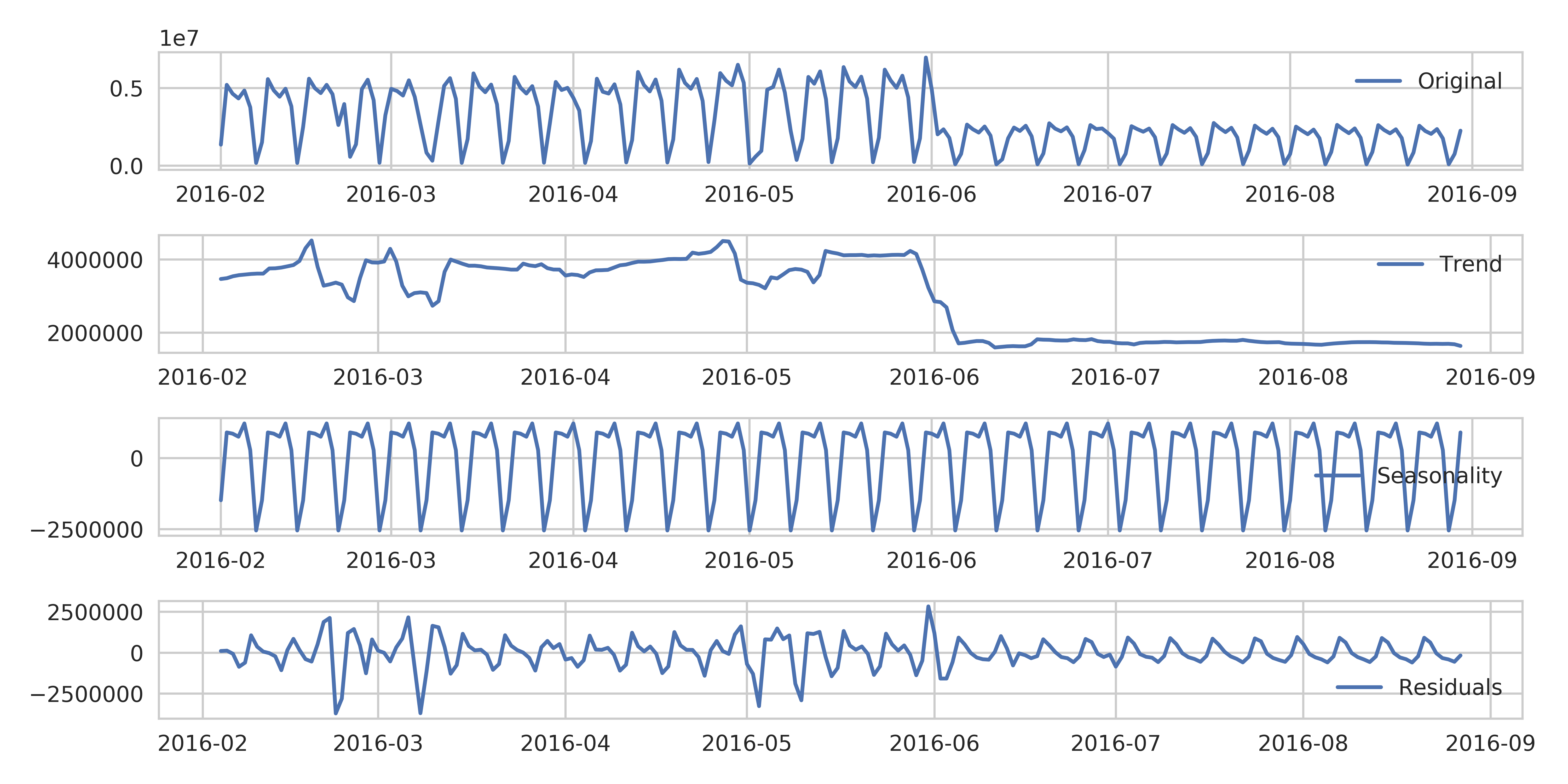}
\caption{Decomposition of daily log demand time series for Fortune 500 company data set}
\label{fig:pmtimeseries}
\end{figure}
Figure \ref{fig:pmtimeseries} exemplifies the stability of the industry by decomposing and plotting the time series of the log demand for 198 days. It is possible to observe a seasonality component and a stable demand. Based on discussions with industry experts, it was decided to discard non-relevant attributes. The reduced data set contained following attributes: (1) 180000 points of sale, (2) 2085 cities, (3) 6 POS types, (4) 39 trade categories, (5) 35 products (SKU), (6) 282 depots. Out of 112 features, the top 5 features according to XGBoost were (1) Revenue by week, SKU and POS, (2) Total volume difference by POS between two consecutive weeks, (3) Total weekly volume by POS, (4) Share of volume by POS at national level, (5) Total revenue difference at depot level between two consecutive weeks. Whereas figure \ref{fig:pmtimeseries} hints at the possibility of using ARIMA or related methods, it would be difficult to obtain significant good results at POS level. The observations by POS were collected on weekly intervals at best and not all SKUs were tracked on each visit. Thus, each POS has at most 28 (7 for Bimbo) points to predict demand for each of the 35 SKUs. This was corroborated during the pre-processing phase. A POS is visited with varying frequency, every 3 to 21 days. On each visit, there could have been notable changes in inventory levels, stocked products and sales volume.
\section{Experiments}\label{ch6}
XGBoost was used on as a baseline for comparisons and to obtain the most relevant features. Additionally, for both data sets a correlation matrix was generated. The observations were fitted onto variations of \textit{GPs} for big data sets such as Parametric \textit{GP} (PGP) \cite{2017arXiv170403144R} and Variational Fourier features for Gaussian processes (VFF) \cite{2016arXiv161106740H}. These two methods were chosen for multiple reasons. First, they represent distinct approaches to \textit{GP} for big data (sparse vs low rank). Second, they are recent approaches in the literature, published in 2017 and 2016 respectively. Third, they can be implemented using GPflow benefiting from GPU acceleration. Fourth, they have shown good results with large data sets.
\paragraph{Hardware}
The models were trained on a server with the operating system Ubuntu Yakkety, a processor Intel Xeon E5-1650 v3 Hexa-Core Haswell, 256 GB DDR4 ECC RAM, a SSD hard drive and a graphic card NVidia GeForce GTX 1080.
\paragraph{Libraries}
For this assessment Pandas 0.20, Numpy 1.12.1, Scikit-learn 0.18, TensorFlow 1.2, GPflow 0.3.8 and XGBoost 0.6a2 were used.
\paragraph{Methodology}
For the purpose of this study, Root Mean Squared Log Error (RMSLE), see definition \ref{ch6:eq1}, was chosen as the evaluation metric.
\begin{equation}\label{ch6:eq1}
RMSLE = \sqrt{\frac{1}{n}\sum_{n=1}^{N}((log(p_i+1)-log(a_i+1))^2}.
\end{equation}
For VFF, an additive Matern$-\frac{3}{2}$ covariance function was chosen, whereas the squared exponential kernel was selected for PGP. This decision was made due to their flexibility and recommendations on the respective literature. Additionally, an automated relevance determination (ARD) kernel was used to identify the best features according to the \textit{GP}.
\begin{figure}[ht]
\centering
  \begin{subfigure}{\linewidth}
  \includegraphics[width=.5\linewidth]{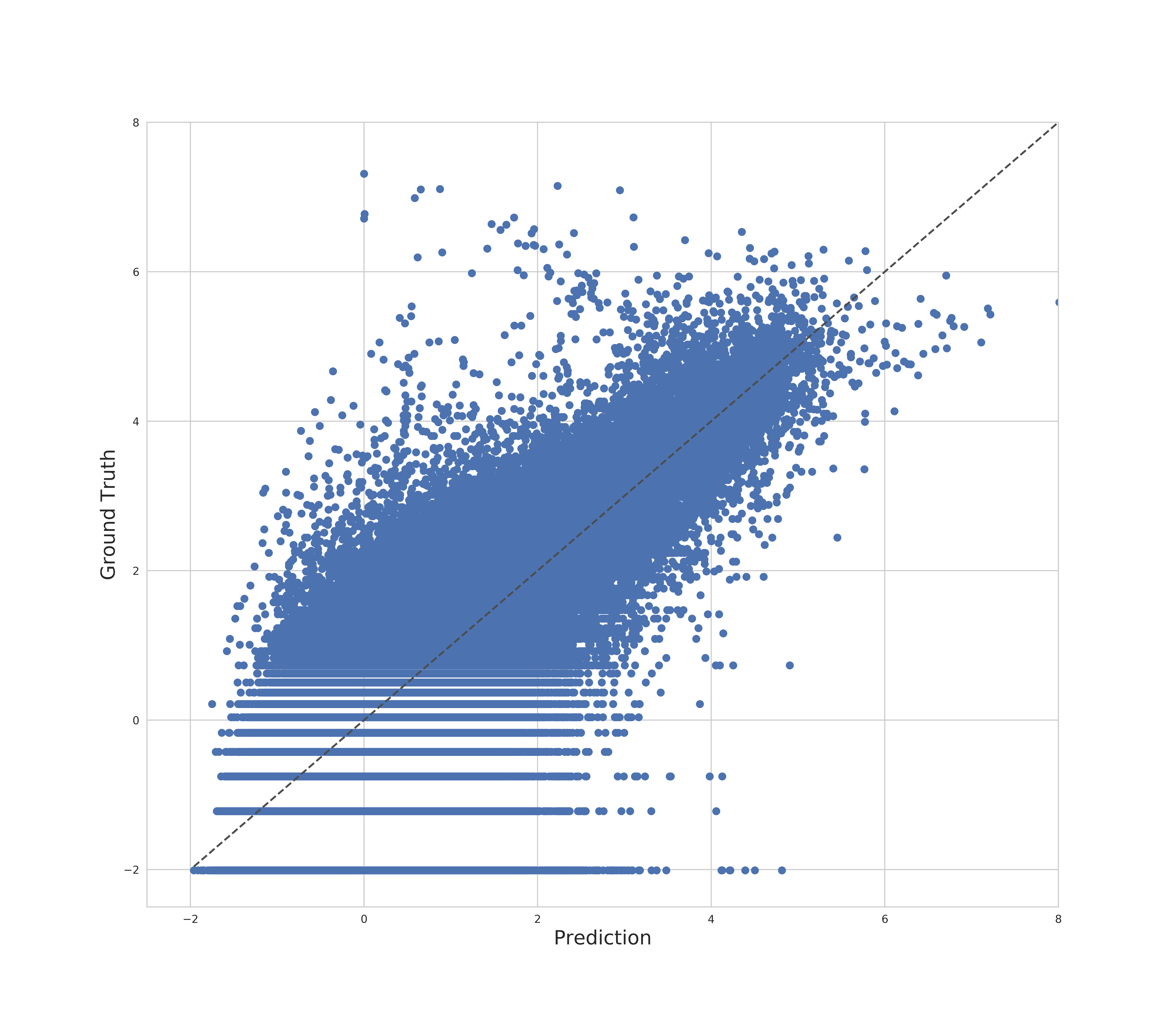}\hfill
  \includegraphics[width=.5\linewidth]{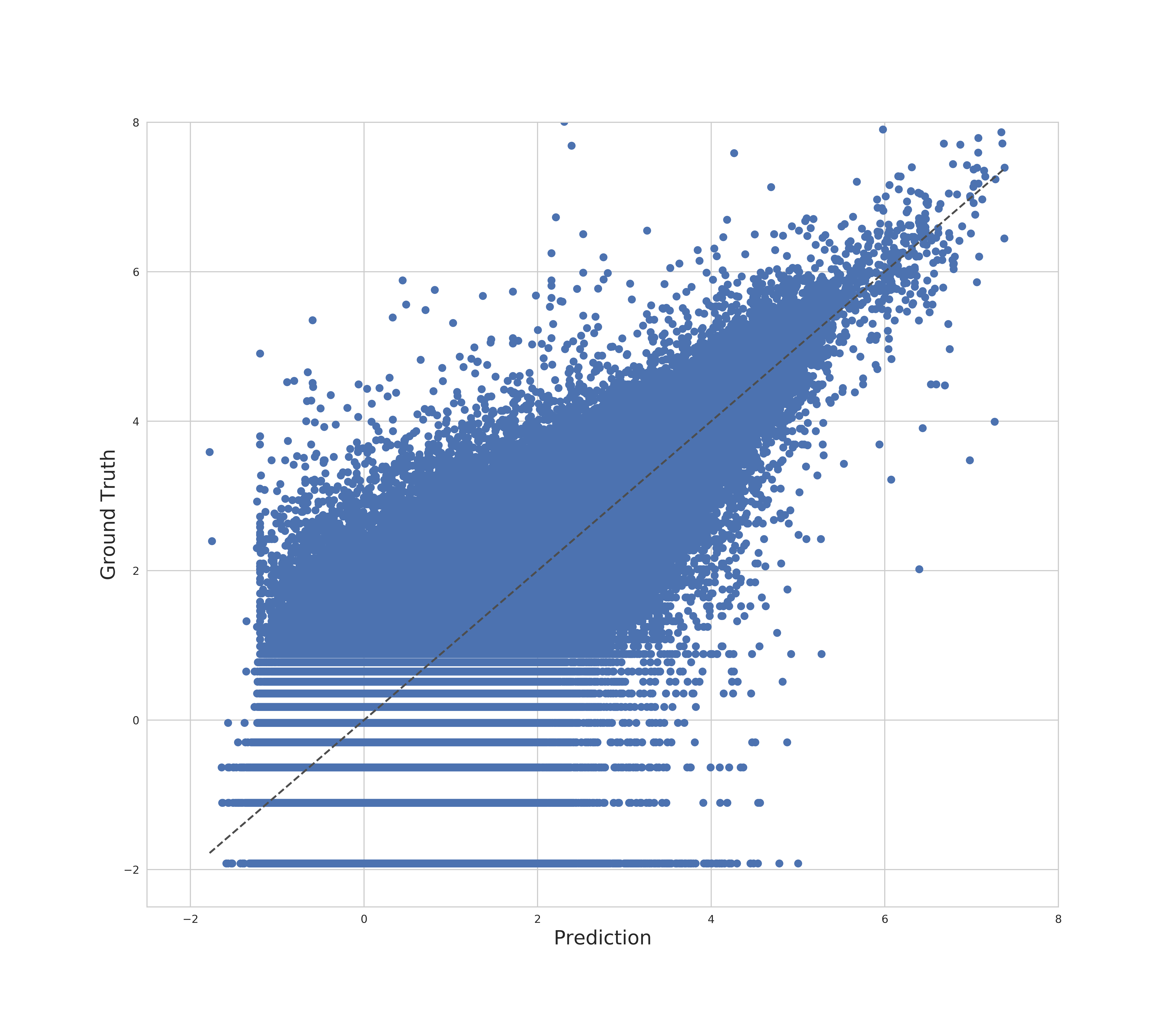}
 \end{subfigure}
\caption{Ground truth (Y-axis) vs prediction (X-axis) for the Bimbo dataset with PGP (left) and VFF (right)}
 \label{fig:bimbogroundvspred}
\end{figure}
\section{Evaluation \& Discussion}\label{ch7}
\begin{figure}
\includegraphics[width=0.5\textwidth]{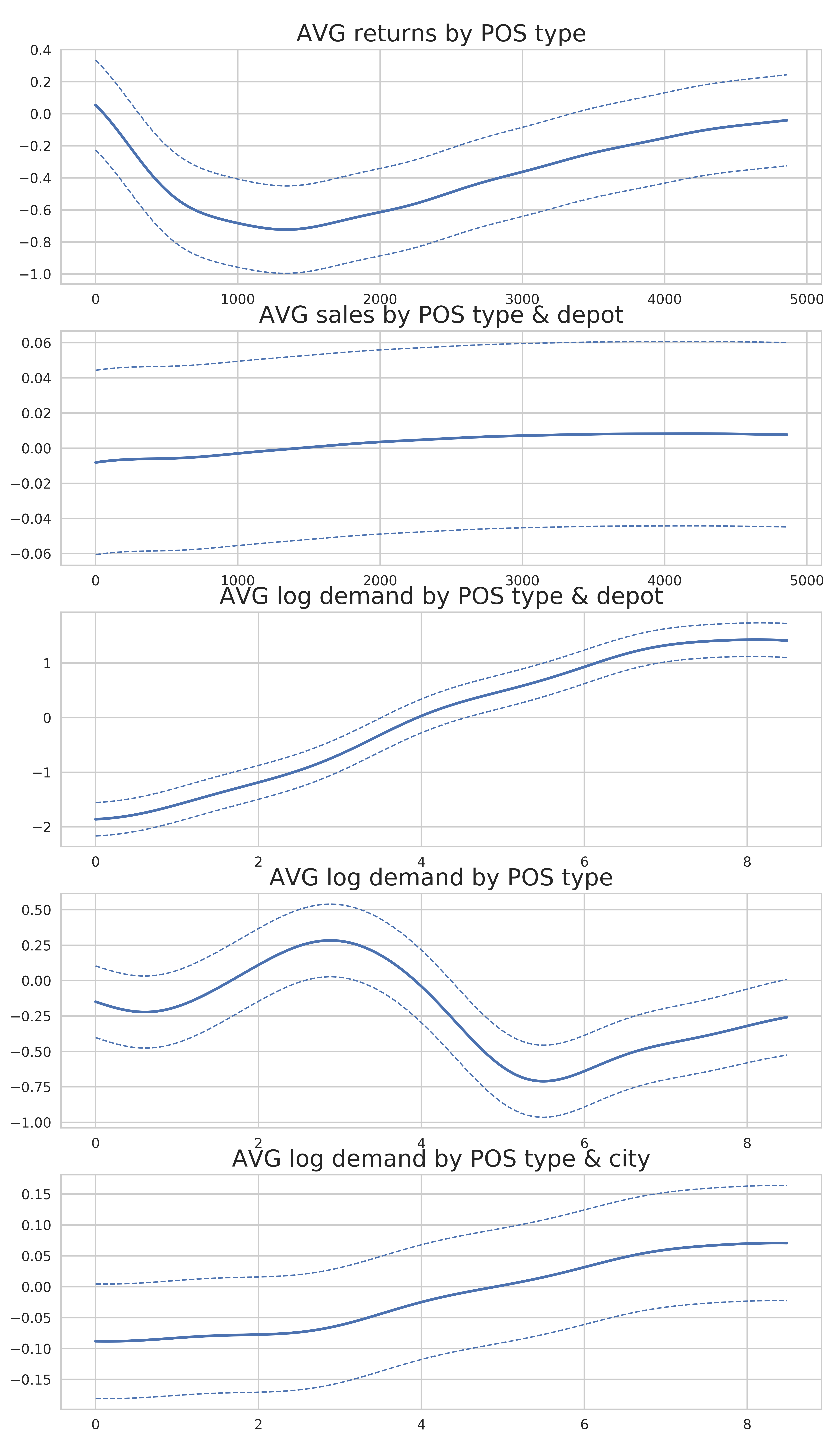}
\caption{Posterior mean of log demand for Top5 XGBoost features for Bimbo data set}
\label{fig:bimboxgbposterior}
\end{figure}

\begin{figure}[ht]
\centering
  \begin{subfigure}{\linewidth}
  \includegraphics[width=.5\linewidth]{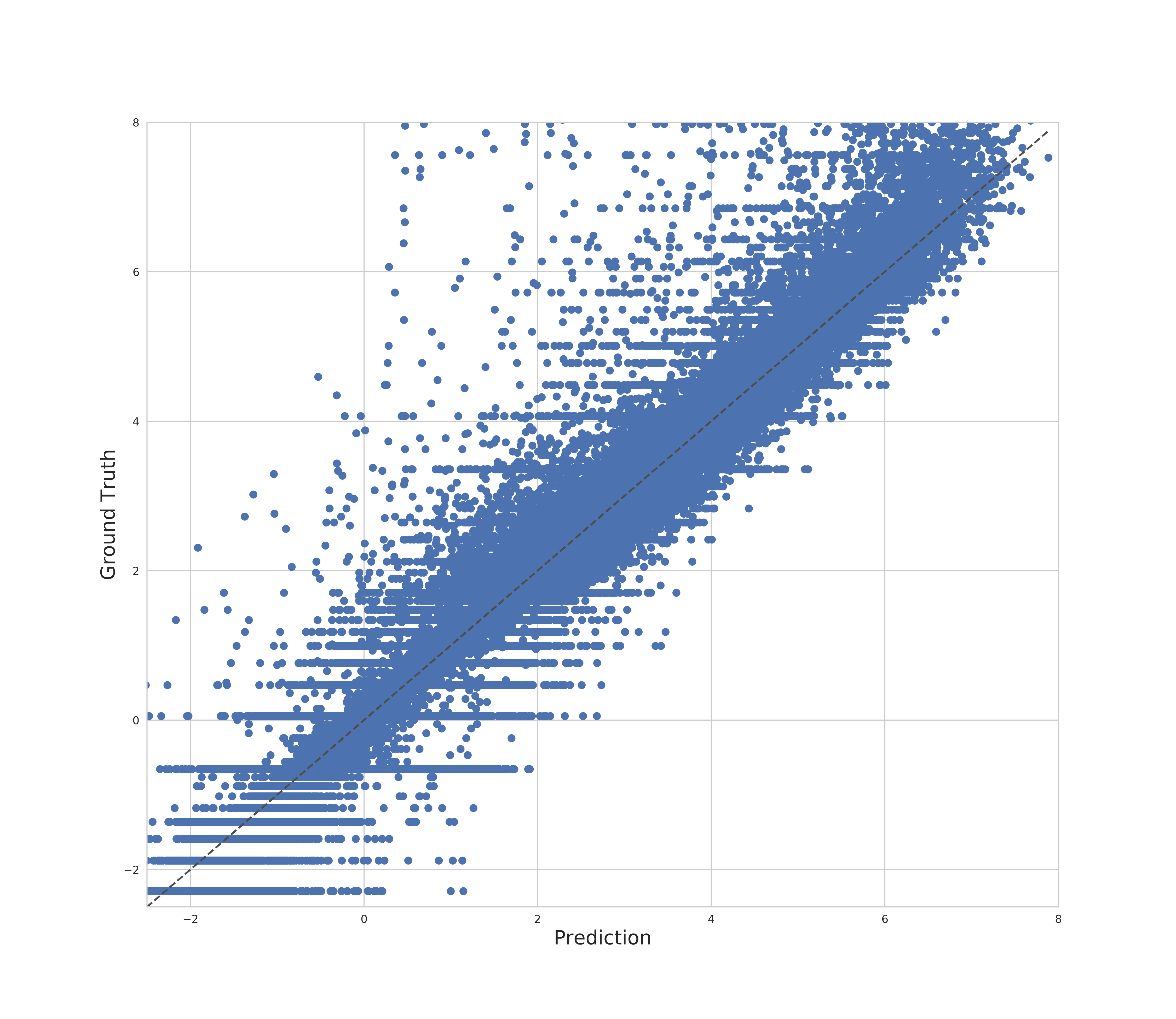}\hfill
  \includegraphics[width=.5\linewidth]{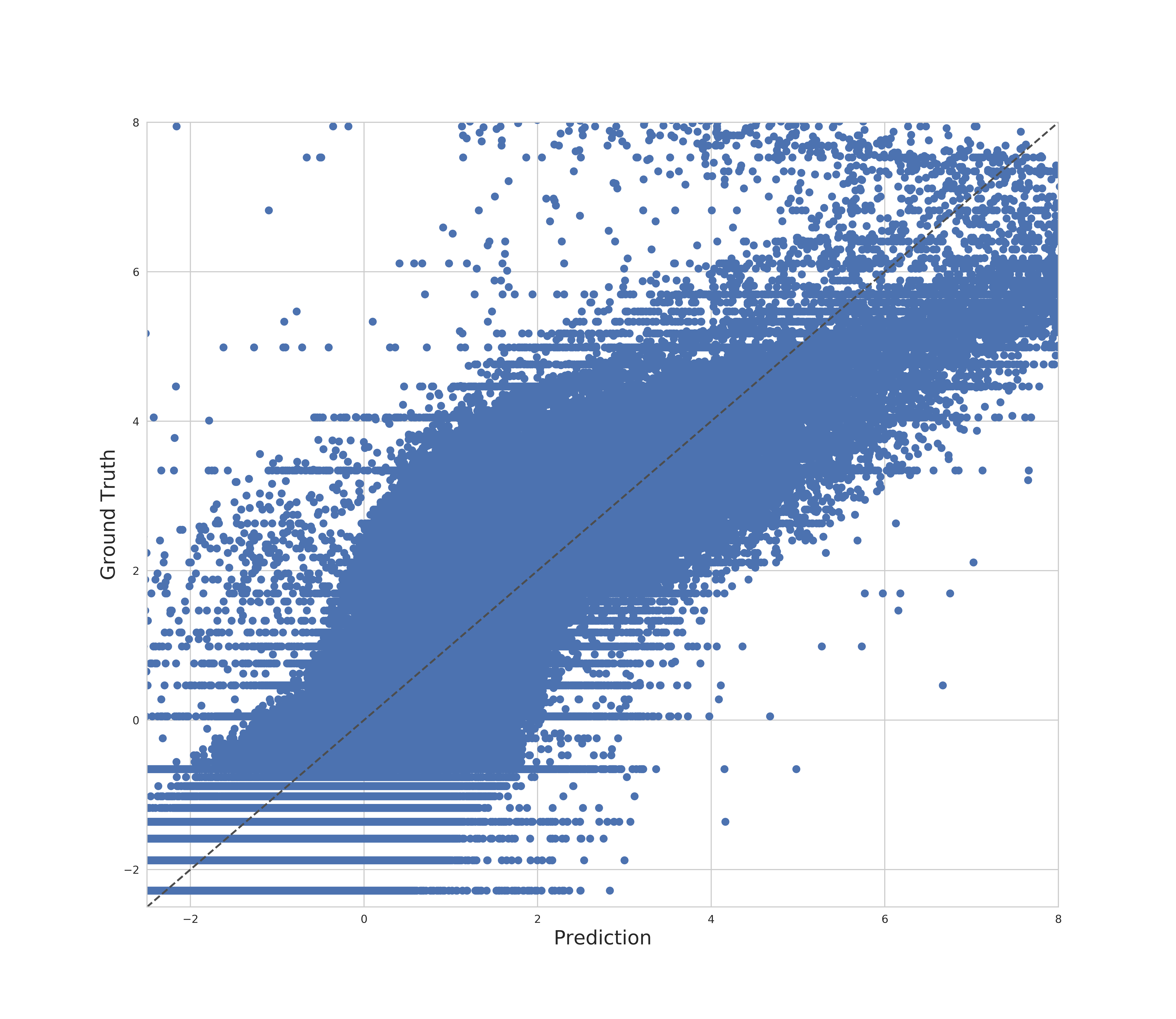}
 \end{subfigure}
\caption{Ground truth (Y-axis) vs prediction (X-axis) for the surveyed company with PGP (left) and VFF (right)}
 \label{fig:pmgroundvspred}
\end{figure}

\begin{figure}
\includegraphics[width=0.5\textwidth]{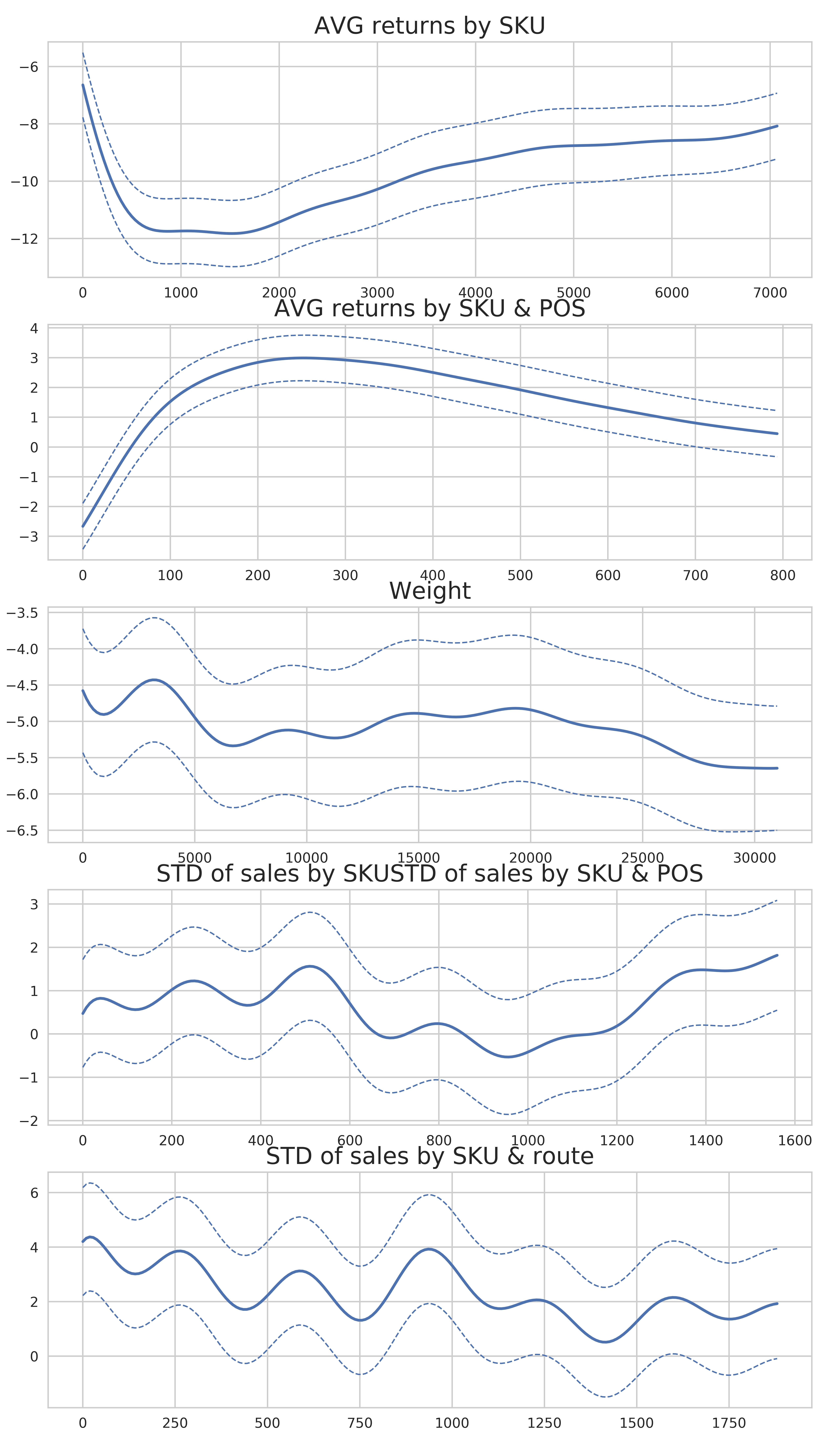}
\caption{Posterior mean of log demand for Top5 ARD features for Bimbo data set}
\label{fig:bimboardposterior}
\end{figure}

For the Bimbo data set, in table \ref{tab:bimboresults}, XGBoost performed better achieving a RMSLE among the top 1\% in the Kaggle competition. On the other hand, PGP was among the top 65\% and VFF was placed among the top 20\% best performers. This can be observed visually in figure \ref{fig:bimbogroundvspred}, where the Y-axis represents the ground truth and the X-axis the prediction. Ideally, the predictive points should be as close as possible to the slope. It is important to highlight that VFF struggled evaluating the full data set. Thus, only the top 10 best features according to XGBoost and to the ARD kernel were used on separate instances. This highlighted the importance of feature selection. The lists of most relevant features according to the correlation matrix, XGBoost and the ARD kernel were very different and none of them can be considered the best one. For the Bimbo data set, the features from XGBoost showed the best results. On a similar vein, once XGBoost was put under the same constraints as VFF and fed only with the top 10 features, performance dropped significantly and fared worse than PGP and VFF. One valuable characteristic of \textit{GPs} is the possibility to evaluate the posterior mean. Decision-makers are often interested in identifying factors that if optimized lead to a certain outcome. For this, the posterior mean can be used. Figure \ref{fig:bimboxgbposterior} depicts the log demand against the Top 5 features according to XGBoost. It can be noticed that the selected features do not have a large impact on log demand. Yet, it helps understand that an increase in returns by POS type impacts log demand, first by decreasing it and later by increasing it. Figure \ref{fig:bimboardposterior} draws similar conclusions depicting average returns by SKU and average returns by SKU and POS. It can be interpreted that returns are a sign of activity. For example, supermarkets have significant sales volume but also large amount of returns. However, once the returns become too large, it signals decreased activity at the POS and thus lower sales. Another insight drawn from evaluating the posterior mean in figure \ref{fig:bimboxgbposterior} is the relation between types of POS (e.g., supermarket, cafeteria, etc) and demand. For example, due to their characteristics, some types of POS generate more demand than others.
For the Fortune 500 company data set, in table \ref{tab:pmresults}, XGBoost achieved a RMSLE of 0.036. The best PGP has a RMSLE of 0.082 obtained in only 15 minutes versus more than 90 minutes for XGBoost. In comparison, VFF showed a RMSLE of 0.44. Figure \ref{fig:pmgroundvspred} plots the ground truth against the predictions for the best models. It is possible to appreciate that PGP has a very low error, whereas VFF struggled. The differences between XGBoost and \textit{GPs} can be attributed to various reasons. On one side, the experiments with VFF used only top features, whereas XGBoost was trained on the full data set. Once XGBoost was trained with the top 10 features only, performance dropped. Similarly, observations with null values had to be discarded to be fitted  the \textit{GP} models. XGBoost, on the other hand, allowed for the presence of null values. Thus, the algorithm had both more and more diverse data at hand. VFF seemed to work better with medium-sized subsets in the range of the hundreds of thousands of observations. Nevertheless, for both \textit{GPs}, it was evident that the choice of covariance functions has a larger impact in the performance over other considerations. This was corroborated when the number of features for VFF was increased from 20 to 50; the results were similar but at a significantly higher time cost. 
\paragraph{Other regression techniques}
Other regression methods such as multivariate adaptive regression splines, e.g., Py-earth\footnote{https://github.com/scikit-learn-contrib/py-earth}, KRR and SVM regression (SVMR) could had been included. However, the main limitations were the lack of production-ready implementations or absence of native libraries. On the other hand, implementations of SVMR and KRR are widely common in Python. However, they proved unsuited for the type of data sets used by this study achieving long training periods and discouraging results.
\paragraph{Other considerations}
A company benefits from a reduction in costs and from added productivity. This can be achieved with the usage of \textit{GP} methods for big data. Although notable for being traditionally slow, new approaches to \textit{GP} turned out to be competitive and even faster than state of the art frameworks such as XGBoost, with the added advantage that GPs did not require a significant effort in parameter tuning. This is replaced with efforts in developing covariance functions. It can be argued that the covariance function is a reflection of the problem task. Thus, the analyst is investing time on understanding the business problem and honing domain knowledge. Similarly, the speed of methods such as PGP allow for quick hypothesis testing and so-called rapid prototyping.
In summary, \textit{GPs} can achieve very similar results to XGBoost under the same conditions, with the added benefit of providing insights to the business on the factors driving demand by analyzing the posterior mean and thus serving as a support tool for decision-makers.
\section{Conclusion}
With two large data sets from the consumer goods sector, it was possible to show that \textit{GPs} can be used as a tool for modeling and prediction for very large data sets. The results are close to state of the art methods such as XGBoost. However, recent frameworks such as PGP proved to be significantly faster and scaled better than XGBoost. An additional benefit is the possibility to analyze the posterior mean conditioned on different features. This points out an additional use case for \textit{GPs} as a decision support tool for management in the \textit{FMCG} sector. In the case of Bimbo, the business can focus its resources on relevant POS with an over-proportional level of returns. This study innovated by providing an overview with depth and breadth on Gaussian Process for big data. It is one of the first academic works on \textit{GPs} for \textit{FMCG} with an additional emphasis on the practitioner. It surveyed more than forty works on \textit{GP} methods for big data. Similarly, it assessed the popularity of Gaussian process among practitioners and benchmarked different \textit{GP} frameworks. Nevertheless, there are still significant areas for innovation. From a theoretical standpoint, a further area of research is to evaluate the development of new kernels considering factors such as seasonality, product life cycle and out of stock events. Similarly, it is necessary to do further work on optimal feature selection for large and high-dimensional data sets. From a practical perspective, there is significant work to assess, which GPR methods for big data are better suited for the \textit{FMCG} industry, and to evaluate the suitability of using the posterior mean as a decision support tool for management. From an implementation point of view, it is interesting to have a unified software framework covering all major GPR methodologies. In conclusion, this work provided an introduction and laid the ground work for future work in \textit{GPs} for \textit{FMCG}. Given the importance of this industry sector in the economy and the benefits of \textit{GPs}, it is to be expected that there will be further academic work in this area.

\begin{table}[ht]
	\caption{Overview of best results for Bimbo data set} \label{tab:bimboresults} 
\centering
	\begin{adjustbox}{max size={\textwidth}{\textheight}}
	\small
\begin{tabular}{| p{1.2cm} | p{1cm} | p{1.5cm} | p{1.5cm} | p{1cm} |}
\hline
Method & Duration & Train set& Test set& RMSLE \\ \hline 
XGBoost & 13.5hr & 5.43M & 3.62M &0.43069 \\ \hline
XGBoost Top10 & 13.5hr & 5.43M & 3.62M & 0.53338 \\ \hline
PGP & 0.25hr & 2.78M & 1.85M& 0.52428 \\ \hline
VFF ARD & 8hr & 4.2M & 2.8M & 0.57227 \\ \hline
VFF XGB & 3.85hr & 60k & 40k & 0.44612 \\ \hline
  \end{tabular}
	\end{adjustbox}
	\end{table}
	
\begin{table}[ht]
	\caption{Overview of best results for surveyed company} \label{tab:pmresults} 
\centering
	\begin{adjustbox}{max size={\textwidth}{\textheight}}
	\small
\begin{tabular}{| p{1.2cm} | p{1cm} | p{1.5cm} | p{1.5cm} | p{1cm} |}
\hline
Method & Duration & Train set& Test set& RMSLE \\ \hline 
XGBoost & 1.37hr & 6.21M & 4.14M & 0.03627 \\ \hline
XGBoost Top10 & 0.83hr & 6.21M & 4.14M & 0.07334 \\ \hline
PGP & 0.83hr & 6.16M & 4.11M & 0.08276 \\ \hline
VFF XGB & 6.33hr & 70k & 30k & 0.44813 \\ \hline
VFF ARD & 7hr & 6.21M & 4.14M & 0.59153 \\ \hline
\end{tabular}
\end{adjustbox}
\end{table}
	
\section{Acknowledgements}
The research, presented in Sections \ref{ch6} and \ref{ch7} of this paper, was supported by the RFBR grants 16-01-00576 A and 16-29-09649 ofi\_m; the research, presented in other sections, was supported by the Russian Science Foundation grant (project 14-50-00150).





%


\bibliography{bibtexDB}

\begin{thebibliography}{10}

\bibitem{tensorflow2016}
Martín Abadi et~al.
\newblock Tensorflow: Large-scale machine learning on heterogeneous distributed
  systems.
\newblock 2016.

\bibitem{hodlr}
S.~{Ambikasaran}, D.~{Foreman-Mackey}, L.~{Greengard}, D.~W. {Hogg}, and
  M.~{O'Neil}.
\newblock {Fast Direct Methods for Gaussian Processes and the Analysis of NASA
  Kepler Mission Data}.
\newblock March 2014.

\bibitem{Banerjee2008}
Sudipto Banerjee, Alan~E Gelfand, and Andrew~O Finley.
\newblock Gaussian predictive process models for large spatial data sets.
\newblock {\em J R Stat Soc Series B Stat Methodol}, 2008.

\bibitem{do2007}
Chuong B.Do.
\newblock Gaussian processes.
\newblock page~13, 2007.

\bibitem{burnaev20162}
M~Belyaev, E~Burnaev, et~al.
\newblock Gtapprox: Surrogate modeling for industrial design.
\newblock In {\em Advances in Engineering Software 102}, pages 29--39, 2016.

\bibitem{burnaev12015}
M~Belyaev, E~Burnaev, and Y~Kapushev.
\newblock Gaussian process regression for structured data sets.
\newblock In {\em Lecture Notes in Artificial Intelligence. Proceedings of
  SLDS}, volume 9047, pages 106--115. Springer, 2015.
\newblock editor:2015. A. Gammerman et al. (Eds.

\bibitem{Blum2013}
Manuel Blum and Martin Riedmiller.
\newblock Electricity demand forecasting using gaussian processes.
\newblock Trading Agent Design and Analysis: Papers from the AAAI 2013
  Workshop, 2013.

\bibitem{burnaev2016}
E~Burnaev, M~Belyaev, and E~Kapushev.
\newblock Computationally efficient algorithm for gaussian processes based
  regression in case of structured samples.
\newblock {\em Computational Mathematics and Mathematical Physics},
  56(4):499--513, 2016.

\bibitem{burnaev20151}
E~Burnaev and M~Panov.
\newblock Adaptive design of experiments based on gaussian processes.
\newblock In {\em Lecture Notes in Artificial Intelligence. Proceedings of
  SLDS}, volume 9047, pages 116--126. Springer, 2015.
\newblock editor:2015. A. Gammerman et al. (Eds.

\bibitem{burnaev20161}
E~Burnaev, M~Panov, and A~Zaytsev.
\newblock Regression on the basis of nonstationary gaussian processes with
  bayesian regularization.
\newblock {\em Journal of Communications Technology and Electronics},
  61(6):661--671, 2016.

\bibitem{burnaev2015}
E~Burnaev and A~Zaytsev.
\newblock Surrogate modeling of mutlifidelity data for large samples.
\newblock {\em Journal of Communications Technology and Electronics},
  60(12):1348--1355, 2015.

\bibitem{9780262182539}
Christopher K.I.~Williams Carl Edward~Rasmussen.
\newblock {\em Gaussian Processes for Machine Learning}.
\newblock MIT Press, 01 2006.

\bibitem{Carpenter_stan}
Bob Carpenter, Daniel Lee, Marcus~A. Brubaker, Allen Riddell, Andrew Gelman,
  Ben Goodrich, Jiqiang Guo, Matt Hoffman, Michael Betancourt, and Peter Li.
\newblock Stan: A probabilistic programming language.

\bibitem{Chase2013}
Charles~Jr Chase.
\newblock {\em Demand-driven forecasting a structured approach to forecasting}.
\newblock Wiley, 2nd edition, 2013.

\bibitem{2014arXiv1408.2060C}
J.~{Chen}, N.~{Cao}, K.~H. {Low}, R.~{Ouyang}, C.~{Keng-Yan Tan}, and
  P.~{Jaillet}.
\newblock {Parallel Gaussian Process Regression with Low-Rank Covariance Matrix
  Approximations}.
\newblock {\em ArXiv e-prints}, August 2014.

\bibitem{Chen2013}
Niya Chen, Zheng Qian, Xiaofeng Meng, and Ian~T Nabney.
\newblock Short-term wind power forecasting using gaussian processes.
\newblock Proceedings of the Twenty-Third International Joint Conference on
  Artificial Intelligence, 2013.

\bibitem{2016arXiv160302754C}
T.~{Chen} and C.~{Guestrin}.
\newblock {XGBoost: A Scalable Tree Boosting System}.
\newblock {\em ArXiv e-prints}, March 2016.

\bibitem{chenren2009}
Tao Chen and Jianghong Ren.
\newblock Bagging for gaussian process regression.
\newblock {\em Neurocomputing}, page~72, 2009.

\bibitem{Claveria2017}
Oscar Claveria, Enric Monte, and Salvador Torra.
\newblock Regional tourism demand forecasting with machine learning models:
  Gaussian process regression vs. neural network models in a multiple-input
  multiple-output setting.
\newblock Technical report, Universitat de Barcelona, Barcelona, 2017.

\bibitem{csatoopper2002}
Lehel Csato and Manfred Opper.
\newblock Sparse online gaussian processes.
\newblock techreport, Aston University, Birmingham B4 7ET, United Kingdom,
  2002.

\bibitem{DBLP:journals/corr/DasRS15}
Sourish Das, Sasanka Roy, and Rajiv Sambasivan.
\newblock Fast gaussian process regression for big data.
\newblock {\em CoRR}, abs/1509.05142, 2015.

\bibitem{2014arXiv1402.4293D}
Alex Davies and Zoubin Ghahramani.
\newblock The random forest kernel and creating other kernels for big data from
  random partitions.
\newblock {\em ArXiv e-prints}, February 2014.

\bibitem{daviesphd2014}
Alexander Davies.
\newblock {\em Effective implementation of gaussian process regression for
  machine learning}.
\newblock phdthesis, University of Cambridge, 2014.

\bibitem{De2016}
Alexander~G De, G~Matthews, Mark Van~Der Wilk, Tom Nickson, Keisuke Fujii,
  Alexis Boukouvalas, Pablo Leon-Villagra, Zoubin Ghahramani, and James
  Hensman.
\newblock Gpflow: A gaussian process library using tensorflow.
\newblock 2016.

\bibitem{Dean:2004:MSD:1251254.1251264}
Jeffrey Dean and Sanjay Ghemawat.
\newblock Mapreduce: Simplified data processing on large clusters.
\newblock In {\em Proceedings of the 6th Conference on Symposium on Opearting
  Systems Design \& Implementation - Volume 6}, OSDI'04, pages 10--10,
  Berkeley, CA, USA, 2004. USENIX Association.

\bibitem{2015arXiv150202843D}
M.~P. {Deisenroth} and J.~W. {Ng}.
\newblock {Distributed Gaussian Processes}.
\newblock {\em ArXiv e-prints}, February 2015.

\bibitem{deisenroth20151}
Marc Deisenroth.
\newblock Distributed gaussian processes.
\newblock page~91. Gaussian Process Summer School, 09.

\bibitem{deisenroth2015}
Marc Deisenroth.
\newblock Gaussian processes for big data problems.
\newblock MLSS, 04 2015.

\bibitem{Dew2016}
Ryan Dew and Asim Ansari.
\newblock Bayesian nonparametric customer base analysis with model-based
  visualizations.
\newblock 2016.

\bibitem{Dew2016a}
Ryan Dew and Asim Ansari.
\newblock Model-based dashboards for customer analytics.
\newblock 2016.

\bibitem{Ding2014}
Fuli Ding and Limin Sun.
\newblock Prediction of tobacco sales based on support vector machine.
\newblock Heidelberg, 2014. Proceedings of 4th International Conference on
  Logistics, Informatics and Service Science, Springer Verlag.

\bibitem{burnaev20171}
Burnaev E. and Zaytsev A.
\newblock Large scale variable fidelity surrogate modeling.
\newblock {\em Annals of Mathematics and Artificial Intelligence}, 2017.

\bibitem{burnaev2017}
Burnaev E and Zaytsev A.
\newblock Minimax approach to variable fidelity data interpolation.
\newblock {\em Proceedings of Machine Learning Research 54:652-661}, 54, 04
  2017.

\bibitem{Erickson_2016}
Collin Erickson, Bruce~E. Ankenman, and Susan~M. Sanchez.
\newblock Comparison of gaussian process modeling software.
\newblock In {\em 2016 Winter Simulation Conference ({WSC})}. {IEEE}, dec 2016.

\bibitem{2017arXiv170309710F}
D.~{Foreman-Mackey}, E.~{Agol}, R.~{Angus}, and S.~{Ambikasaran}.
\newblock {Fast and scalable Gaussian process modeling with applications to
  astronomical time series}.
\newblock {\em ArXiv e-prints}, March 2017.

\bibitem{furrer2006}
R~Furrer, M.~G Genton, and D~Nychka.
\newblock Covariance tapering for interpolation of large spatial datasets.
\newblock {\em Journal of Computational and Graphical Statistics},
  15(3):502–523, 2006.

\bibitem{2014arXiv1402.1389G}
Yarin Gal, Mark Van~Der Wilk, and Carl~E Rasmussen.
\newblock {Distributed Variational Inference in Sparse Gaussian Process
  Regression and Latent Variable Models}.
\newblock {\em ArXiv e-prints}, February 2014.

\bibitem{gilboa2013}
Elad Gilboa, Yunus Saatci, and John~P. Cunningham.
\newblock Scaling multidimensional gaussian processes using projected additive
  approximations.
\newblock volume Proceedings of the 30th International Conference on Machine
  Learning, Atlanta, Georgia, USA, 2013. JMLR.

\bibitem{CIS-180533}
Tilmann Gneiting.
\newblock Compactly supported correlation functions.
\newblock {\em Journal of Multivariate Analysis}, 83(2):493--508, 2002.

\bibitem{erogol2016}
Eren Golge.
\newblock Comparison of deep learning libraries after years of use.
\newblock Internet, April 2016.

\bibitem{gpy2014}
{GPy}.
\newblock {GPy}: A gaussian process framework in python, 2012.

\bibitem{2013arXiv1303.0383G}
R.~B. {Gramacy} and D.~W. {Apley}.
\newblock {Local Gaussian process approximation for large computer
  experiments}.
\newblock {\em ArXiv e-prints}, March 2013.

\bibitem{2013arXiv1310.5182G}
R.~B. {Gramacy}, J.~{Niemi}, and R.~M. {Weiss}.
\newblock {Massively parallel approximate Gaussian process regression}.
\newblock {\em ArXiv e-prints}, October 2013.

\bibitem{Arxiv0710.4536}
Robert~B. Gramacy and Herbert K.~H. Lee.
\newblock Bayesian treed gaussian process models with an application to
  computer modeling.
\newblock {\em ArXiv e-prints}, 2009.

\bibitem{2016arXiv161008035G}
A.~{Grigorievskiy}, N.~{Lawrence}, and S.~{S{\"a}rkk{\"a}}.
\newblock {Parallelizable sparse inverse formulation Gaussian processes
  (SpInGP)}.
\newblock {\em ArXiv e-prints}, October 2016.

\bibitem{2016arXiv161106740H}
J.~{Hensman}, N.~{Durrande}, and A.~{Solin}.
\newblock {Variational Fourier features for Gaussian processes}.
\newblock {\em ArXiv e-prints}, November 2016.

\bibitem{DBLP:conf/uai/2013}
James Hensman, Nicol{\'{o}} Fusi, and Neil~D. Lawrence.
\newblock Gaussian processes for big data.
\newblock In {\em Proceedings of the Twenty-Ninth Conference on Uncertainty in
  Artificial Intelligence, {UAI} 2013, Bellevue, WA, USA, August 11-15, 2013},
  2013.

\bibitem{2016arXiv1611.06080}
Quang~Minh Hoang, Trong~Nghia Hoang, and Kian~Hsiang Low.
\newblock A generalized stochastic variational bayesian hyperparameter learning
  framework for sparse spectrum gaussian process regression.
\newblock 2016.

\bibitem{hoang2014}
Trong~Nghia Hoang.
\newblock {\em New Advances on Bayesian and Decision-Theoretic Approaches for
  Interactive Machine Learning}.
\newblock phdthesis, National University of Singapore, 2014.

\bibitem{pmlr-v37-hoang15}
Trong~Nghia Hoang, Quang~Minh Hoang, and Bryan Kian~Hsiang Low.
\newblock A unifying framework of anytime sparse gaussian process regression
  models with stochastic variational inference for big data.
\newblock In Francis Bach and David Blei, editors, {\em Proceedings of the 32nd
  International Conference on Machine Learning}, volume~37 of {\em Proceedings
  of Machine Learning Research}, pages 569--578, Lille, France, 07--09 Jul
  2015. PMLR.

\bibitem{hoang206}
Trong~Nghia Hoang, Quang~Minh Hoang, and Bryan Kian~Hsiang Low.
\newblock A distributed variational inference framework for unifying parallel
  sparse gaussian process regression models.
\newblock In {\em Proceedings of the 33rd International Conference on Machine
  Learning}, number~48, New York, NY, USA, 2016. JLMR.

\bibitem{Huang2015}
Wenjie Huang, Qing Zhang, Wei Xu, Hongjiao Fu, Mingming Wang, and Xun Liang.
\newblock A novel trigger model for sales prediction with data mining
  techniques.
\newblock 2015.

\bibitem{Kaufman08covariancetapering}
Cari Kaufman.
\newblock Covariance tapering for likelihoodbased estimation in large spatial
  data sets.
\newblock {\em Journal of the American Statistical Association}, pages
  1545--1555, 2008.

\bibitem{Kosar2016}
Arda Kosar, Hayes Cozart, and Kyle Szela.
\newblock Predicting demand from historical sales data- grupo bimbo kaggle
  competition, 2016.

\bibitem{Lawrence2016}
Neil Lawrence.
\newblock Introduction to gaussian processes.
\newblock page 504. MLSS, 08 2016.

\bibitem{Lindgren_2011}
Finn Lindgren, H{\aa}vard Rue, and Johan Lindström.
\newblock An explicit link between gaussian fields and gaussian markov random
  fields: the stochastic partial differential equation approach.
\newblock {\em Journal of the Royal Statistical Society: Series B (Statistical
  Methodology)}, 73(4):423--498, aug 2011.

\bibitem{2014arXiv1411.4510L}
K.~H. {Low}, J.~{Yu}, J.~{Chen}, and P.~{Jaillet}.
\newblock {Parallel Gaussian Process Regression for Big Data: Low-Rank
  Representation Meets Markov Approximation}.
\newblock {\em ArXiv e-prints}, November 2014.

\bibitem{Lu2012}
Chi~Jie Lu, Tian~Shyug Lee, and Chia~Mei Lian.
\newblock Sales forecasting for computer wholesalers: A comparison of
  multivariate adaptive regression splines and artificial neural networks.
\newblock {\em Decision Support Systems}, 2012.

\bibitem{gpfit2013}
Blake MacDonald, Pritam Ranjan, and Hugh Chipman.
\newblock Gpfit: An r package for gaussian process model fitting using a new
  optimization algorithm.
\newblock {\em ArXiv e-prints}, 2013.

\bibitem{Neumann:2015:PPL:2789272.2912082}
Marion Neumann, Shan Huang, Daniel~E. Marthaler, and Kristian Kersting.
\newblock pygps: A python library for gaussian process regression and
  classification.
\newblock {\em J. Mach. Learn. Res.}, 16(1):2611--2616, January 2015.

\bibitem{Nguyen_Tuong_2009}
Duy Nguyen-Tuong, Matthias Seeger, and Jan Peters.
\newblock Model learning with local gaussian process regression.
\newblock {\em Advanced Robotics}, 23(15):2015--2034, jan 2009.

\bibitem{2017arXiv170106655P}
Chiwoo Park and Daniel~W. Apley.
\newblock {Patchwork Kriging for Large-scale Gaussian Process Regression}.
\newblock {\em CoRR}, abs/1701.06655, January 2017.

\bibitem{JMLR:v17:15-327}
Chiwoo Park and Jianhua~Z. Huang.
\newblock Efficient computation of gaussian process regression for large
  spatial data sets by patching local gaussian processes.
\newblock {\em Journal of Machine Learning Research}, 17(174):1--29, 2016.

\bibitem{Park:2011:DDA:1953048.2021054}
Chiwoo Park, Jianhua~Z. Huang, and Yu~Ding.
\newblock Domain decomposition approach for fast gaussian process regression of
  large spatial data sets.
\newblock {\em J. Mach. Learn. Res.}, 12:1697--1728, July 2011.

\bibitem{FABIANPEDREGOSA2011}
Fabian Pedregosa, Alexandre Gramfort, and Vincent Michel.
\newblock Scikit-learn: Machine learning in python.
\newblock {\em Journal of Machine Learning Research}, 12:2825--2830, 2011.

\bibitem{2745}
J.~Quinonero~Candela and CE. Rasmussen.
\newblock Analysis of some methods for reduced rank gaussian process
  regression.
\newblock In {\em Switching and Learning in Feedback Systems}, pages 98--127,
  Berlin, Germany, 2005. Max-Planck-Gesellschaft, Springer.

\bibitem{4798}
J.~Quiñonero-Candela, CE. Rasmussen, and CKI. Williams.
\newblock {\em Approximation Methods for Gaussian Process Regression}, pages
  203--223.
\newblock Neural Information Processing. MIT Press, Cambridge, MA, USA,
  September 2007.

\bibitem{2017arXiv170403144R}
M.~{Raissi}.
\newblock {Parametric Gaussian Process Regression for Big Data}.
\newblock {\em ArXiv e-prints}, April 2017.

\bibitem{Rasmussen2006}
Carl~Edward Rasmussen.
\newblock Gaussian processes in machine learning.
\newblock page~9, 2006.

\bibitem{Rasmussen01infinitemixtures}
Carl~Edward Rasmussen and Zoubin Ghahramani.
\newblock Infinite mixtures of gaussian process experts.
\newblock In {\em In Advances in Neural Information Processing Systems 14},
  pages 881--888. MIT Press, 2001.

\bibitem{2017arXiv170305687R}
R.~R. {Richardson}, M.~A. {Osborne}, and D.~A. {Howey}.
\newblock {Gaussian process regression for forecasting battery state of
  health}.
\newblock {\em ArXiv e-prints}, March 2017.

\bibitem{Salvatier2016}
John Salvatier, Thomas~V Wiecki, and Christopher Fonnesbeck.
\newblock Probabilistic programming in python using pymc3.
\newblock {\em PeerJ Computer Science}, 2:55, 04 2016.

\bibitem{Samarasinghe2012}
Milindanath Samarasinghe, Waseem Al, Hawani, and Ole-Christoffer Granmo.
\newblock Short-term forecasting of electricity consumption using gaussian
  processes.
\newblock Technical report, 2012.

\bibitem{Seeger2004}
Matthias Seeger.
\newblock Gaussian processes for machine learning.
\newblock 2004.

\bibitem{Seeger03fastforward}
Matthias Seeger, Christopher K.~I. Williams, and Neil~D. Lawrence.
\newblock Fast forward selection to speed up sparse gaussian process
  regression.
\newblock In {\em IN WORKSHOP ON AI AND STATISTICS 9}, 2003.

\bibitem{Snelson07localand}
Edward Snelson.
\newblock Local and global sparse gaussian process approximations.
\newblock In {\em Proceedings of Artificial Intelligence and Statistics
  (AISTATS}, 2007.

\bibitem{NIPS2005_2857}
Edward Snelson and Zoubin Ghahramani.
\newblock Sparse gaussian processes using pseudo-inputs.
\newblock In Y.~Weiss, P.~B. Sch\"{o}lkopf, and J.~C. Platt, editors, {\em
  Advances in Neural Information Processing Systems 18}, pages 1257--1264. MIT
  Press, 2006.

\bibitem{Tan2015}
Jian Tan.
\newblock Guizhou cigarette sales prediction based on seasonal decomposition
  mlp.
\newblock International Symposium on Social Sciences (ISSS 20015), 2015.

\bibitem{Titsias09variationallearning}
Michalis~K. Titsias.
\newblock Variational learning of inducing variables in sparse gaussian
  processes.
\newblock In {\em In Artificial Intelligence and Statistics 12}, pages
  567--574, 2009.

\bibitem{Tran2017}
Dustin Tran, Alp Kucukelbir, et~al.
\newblock Edward: A library for probabilistic modeling, inference, and
  criticism.
\newblock 2017.

\bibitem{Tresp_2000}
Volker Tresp.
\newblock A bayesian committee machine.
\newblock {\em Neural Computation}, 12(11):2719--2741, nov 2000.

\bibitem{conf/cvpr/UrtasunD08}
Raquel Urtasun and Trevor Darrell.
\newblock Sparse probabilistic regression for activity-independent human pose
  inference.
\newblock In {\em CVPR}. IEEE Computer Society, 2008.

\bibitem{2012arXiv1206.3290V}
J.~{Vanhatalo} and A.~{Vehtari}.
\newblock {Modelling local and global phenomena with sparse Gaussian
  processes}.
\newblock {\em ArXiv e-prints}, June 2012.

\bibitem{Wang2014}
Ye~Wang, Carlos Ocampo-Martínez, et~al.
\newblock Gaussian-process-based demand forecasting for predictive control of
  drinking water networks.
\newblock 2014.

\bibitem{Wu2012}
Qi~Wu, Rob Law, and Xin Xu.
\newblock A sparse gaussian process regression model for tourism demand
  forecasting in hong kong.
\newblock {\em Expert Systems with Applications}, 2012.

\bibitem{Zaharia:2010:SCC:1863103.1863113}
Matei Zaharia, Mosharaf Chowdhury, Michael~J. Franklin, Scott Shenker, and Ion
  Stoica.
\newblock Spark: Cluster computing with working sets.
\newblock In {\em Proceedings of the 2Nd USENIX Conference on Hot Topics in
  Cloud Computing}, HotCloud'10, pages 10--10, Berkeley, CA, USA, 2010. USENIX
  Association.

\bibitem{burnaev2013}
A~Zaitsev, E~Burnaev, and V~Spokoiny.
\newblock Properties of the posterior distribution of a regression model based
  on gaussian random fields.
\newblock {\em Automation and Remote Control}, 74(10):1645--1655, 2013.

\bibitem{Zhang2017}
Michael~M Zhang and Sinead~A Williamson.
\newblock Embarrassingly parallel inference for gaussian processes.
\newblock {\em ArXiv e-prints}, February 2017.

\end{thebibliography}
\bibliographystyle{plain}

\end{document}